\documentclass[aps,prl,twocolumn,showpacs,superscriptaddress,nofootinbib,preprintnumbers]{revtex4-1}  
\usepackage{graphicx,color}  
\usepackage{dcolumn}   
\usepackage{bm}        
\usepackage{amssymb} 
\usepackage{amsmath}
 \usepackage{slashed}
 \usepackage{multirow}
\usepackage{amsmath,amsfonts,amstext,amssymb,mathrsfs}

\def\){\right)}
\def\({\left( }
\def\]{\right] }
\def\[{\left[ }

\newcommand{\be}{\begin{equation}}
\newcommand{\ee}{\end{equation}}

\def\bea{\begin{eqnarray}}
\def\eea{\end{eqnarray}}

\def\bal#1\eal{\begin{align}#1\end{align}}

\def\bald{\begin{aligned}}
\def\eald{\end{aligned}}

\def\bsub{\begin{subequations}}
\def\esub{\end{subequations}}

\def\beqx{\begin{displaymath}}
\def\eeqx{\end{displaymath}}

\newcommand{\bmat}{\left(\begin{array}}
\newcommand{\emat}{\end{array}\right)}

\def\a{\alpha}

\def\c{\chi}
\def\d{\delta}
\def\e{\epsilon}
\def\f{\phi}
\def\g{\gamma}
\def\h{\eta}
\def\j{\psi}
\def\k{\kappa}
\def\l{\lambda}
\def\m{\mu}
\def\n{\nu}
\def\o{\omega}
    
\def\p{\pi}

    \def\th{\theta}
\def\r{\rho}
\def\s{\sigma}

\def\x{\xi}

\def\G{\Gamma}

\def\L{\Lambda}

\def\ve{\varepsilon}

\def\ca{{\cal A}}

\def\cj{{\cal J}}

\def\cn{{\cal N}}
\def\co{{\cal O}}
\def\cp{{\cal P}}
\def\cq{{\cal Q}}

\def\ct{{\cal T}}

\def\bo{{\raise-.3ex\hbox{\large$\Box$}}}               
\def\pa{\partial}                                       
\def\face{{\raise.2ex\hbox{$\displaystyle \bigodot$}\mskip-2.2mu \llap {$\ddot
        \smile$}}}                                   
\def\>{\rangle}                                      
\def\<{\langle}                                      

\def\wt#1{\widetilde{#1}}                            
\def\lbar#1{\ensuremath{\overline{#1}}}              
\def\leftrightarrowfill{$\mathsurround=0pt \mathord\leftarrow \mkern-6mu
        \cleaders\hbox{$\mkern-2mu \mathord- \mkern-2mu$}\hfill
        \mkern-6mu \mathord\rightarrow$}        
\def\dvec#1{\vbox{\ialign{##\crcr
        \leftrightarrowfill\crcr\noalign{\kern-1pt\nointerlineskip}
        $\hfil\displaystyle{#1}\hfil$\crcr}}}           
\def\Re{{\rm Re\,}}                                     
\def\Im{{\rm Im\,}}                                     

\def\-{\hphantom{-}}

\begin{document}

\def\eg{{\it e.g.}}
\newcommand{\nc}{\newcommand}
\nc{\rnc}{\renewcommand}
\nc{\K}{\kappa}
\nc{\bK}{\bar{\K}}
\nc{\bN}{\bar{N}}
\nc{\bq}{\bar{q}}
\nc{\vbq}{\vec{\bar{q}}}
\nc{\lrarrow}{\leftrightarrow}
\nc{\rg}{\sqrt{g}}
\rnc{\[}{\begin{equation}}
\rnc{\]}{\end{equation}}
\nc{\nn}{\nonumber}
\rnc{\(}{\left(}
\rnc{\)}{\right)}
\nc{\ep}{\epsilon}
\nc{\tto}{\rightarrow}
\rnc{\inf}{\infty}
\rnc{\Re}{\mathrm{Re}}
\rnc{\Im}{\mathrm{Im}}
\nc{\mA}{\mathcal{A}}
\nc{\mB}{\mathcal{B}}
\nc{\mC}{\mathcal{C}}
\nc{\mD}{\mathcal{D}}
\nc{\mN}{\mathcal{N}}
\rnc{\H}{\mathcal{H}}
\rnc{\L}{\mathcal{L}}
\rnc{\>}{\rangle}
\nc{\fnl}{f_{NL}}
\nc{\gnl}{g_{NL}}
\nc{\fnleq}{f_{NL}^{equil.}}
\nc{\fnlloc}{f_{NL}^{local}}
\nc{\Lie}{\pounds}
\nc{\bOmega}{\bar{\Omega}}
\nc{\bLambda}{\bar{\Lambda}}
\nc{\dN}{\delta N}
\nc{\gYM}{g_{\mathrm{YM}}}
\nc{\geff}{g_{\mathrm{eff}}}
\nc{\oa}{\stackrel{\leftrightarrow}}
\nc{\IR}{{\rm IR}}
\nc{\UV}{{\rm UV}}

\title{Anomalous supersymmetry} 

\author{Georgios Katsianis}  \affiliation{STAG Research Centre, Highfield, University of Southampton, SO17 1BJ Southampton, UK}  \affiliation{Mathematical Sciences, Highfield, University of Southampton, SO17 1BJ Southampton, UK}
\author{Ioannis Papadimitriou} \affiliation{School of Physics, Korea Institute for Advanced Study, Seoul 02455, Korea} 
\author{Kostas Skenderis} \affiliation{STAG Research Centre, Highfield, University of Southampton, SO17 1BJ Southampton, UK}  \affiliation{Mathematical Sciences, Highfield, University of Southampton, SO17 1BJ Southampton, UK}
 \author{Marika Taylor} \affiliation{STAG Research Centre, Highfield, University of Southampton, SO17 1BJ Southampton, UK}  \affiliation{Mathematical Sciences, Highfield, University of Southampton, SO17 1BJ Southampton, UK}

\date{\today}

\preprint{KIAS-P19008}

\begin{abstract}
We show that supersymmetry is anomalous in ${\cal N}=1$ superconformal quantum field theories (SCFTs) with an anomalous R-symmetry. This anomaly was originally found in holographic SCFTs at strong coupling. Here we show that this anomaly is present in general and demonstrate it for the massless superconformal Wess-Zumino model via an one loop computation. The anomaly appears first in four-point functions of two supercurrents with either two R-currents or with an R-current and an energy momentum tensor. In fact, the Wess-Zumino consistency conditions together with the standard R-symmetry anomaly imply the existence of the anomaly. We outline the implications of this anomaly. 

\end{abstract}

\maketitle


Anomalies of symmetries play an important role in quantum field theories. If a global symmetry is anomalous, classical selection rules are not respected in the quantum theory and classically forbidden processes may occur. This is a feature of the theory and it is linked with observable effects. For example, the axial anomaly explains the $\pi^0$ decay and leads to the resolution of the $U(1)$ problem in QCD \cite{Adler:1969gk,Bell:1969ts}. On the other hand, anomalies in local (gauge) symmetries lead to inconsistencies, such as lack of unitarity,  and they must be canceled. An important  corollary is that anomalous global symmetries cannot be consistently coupled to corresponding local symmetries.
Reviews on anomalies in quantum field theories may be found in \cite{Treiman:1986ep,Fujikawa:2004cx}.


{\em Anomalies in supersymmetric theories.}--- In this paper we discuss a new anomaly
 in four-dimensional supersymmetric quantum field theories with an anomalous R-symmetry: 
 global supersymmetry itself is anomalous. This anomaly was discovered in the context of superconformal theories that can be realised holographically \cite{Papadimitriou:2017kzw}. Here we show that the same anomaly arises in perturbation theory in the simplest supersymmetric model: the free superconformal Wess-Zumino (WZ) model.  
 
 An anomaly may be detected either by putting the theory on a non-trivial background, or by computing correlation functions on a flat background and checking whether the Ward identities are satisfied. The latter method was the one that led to the original discovery of anomalies via one-loop triangle diagrams \cite{Adler:1969gk,Bell:1969ts}. Here we will carry out the analogous computation for the supersymmetry anomaly. The anomaly is associated in particular with anomalous one-loop contributions to four-point correlation functions between two supersymmetry currents and
 two R-currents or an R-current and an energy momentum tensor. We will discuss the former in the free superconformal WZ model but analogous contributions would arise in any supersymmetric theory with a (softly broken) anomalous R-symmetry. Actually, as will be sketched below and is shown in detail in the companion paper \cite{Papadimitriou:2019gel}, the WZ consistency conditions \cite{Wess:1971yu} together with the standard triangle anomalies imply that supersymmetry must be anomalous.

Discussion of anomalies in 4d (super)conformal QFT has a long history. It has been known since the 1970s \cite{Capper:1974ic,Deser:1976yx} that the trace of the stress tensor ${\cal T}^{\mu}_{\mu}$ is anomalous in the presence of a curved background metric $g_{\mu \nu}$ 
and background source $A_{\mu}$ for a chiral current ${\cal J}_\mu$, and 
the R-current is  similarly anomalous. Moreover, there are generally mixed anomalies involving 
two energy momentum tensors and a chiral current \cite{Delbourgo:1972xb, AlvarezGaume:1983ig}.
It has also been known since \cite{Ferrara:1974pz} that the currents sit in a supermultiplet, as do the anomalies. In particular, the trace anomaly and the R-current anomaly are in the same multiplet as the 
gamma trace of the supercurrent, $\gamma^\mu {\cal Q}_\mu$. The latter is an anomaly in the conservation of the special supersymmetry current, $x^\nu \gamma_{\nu}  {\cal Q}_\mu$. 
It follows that special supersymmetry (sometimes also called S-supersymmetry) is anomalous. It was believed however that supersymmetry itself (sometimes called Q-supersymmetry) is preserved, i.e. the conservation of ${\cal Q}_\mu$ is non-anomalous. 

There have been extensive studies in the past regarding anomalies in supersymmetry.
It was realised early on \cite{deWit:1975veh,Abbott:1977in,Abbott:1977xj,Abbott:1977xk,Hieda:2017sqq,Batista:2018zxf} that one cannot maintain at the quantum level simultaneously  $\partial^\mu {\cal Q}_\mu=0$
and $\gamma^\mu  {\cal Q}_\mu=0$ and, if the model is a gauge theory, gauge invariance: one of the three conditions must be relaxed and the standard choice is to have a superconformal anomaly.
This is the standard superconformal anomaly mentioned above and is distinct from the anomaly discussed here. Also distinct is the Konishi anomaly \cite{Konishi:1983hf, Konishi:1985tu}, which is a superspace version of the chiral anomaly in supersymmetric gauge theories.

Another set of studies, reviewed in \cite{Piguet:1986ug}, considers the effective action for elementary fields and examines whether it is invariant under supersymmetry including loop effects; it investigates  the conservation of the supercurrent inside correlators of elementary fields and/or solves the WZ consistency conditions relevant for this setup, and finds no supersymmetry anomaly. This does not contradict the results we present below: to find the anomaly one should either put the theory on a non-trivial background or consider correlation functions of (classically) conserved currents\footnote{To illustrate this point, consider a free fermion in a complex representation in flat spacetime. This theory has a standard axial anomaly originating from the 3-point function of the axial current. However, if one only looks at correlators of elementary fields these are non-anomalous  and the axial current inside such correlators is conserved.}. Studies involving correlators of currents have also appeared but typically only discuss 3-point functions of bosonic currents.
As mentioned above, the supersymmetry anomaly appears first in 4-point functions involving two supercurrents and two bosonic currents and to our knowledge these have not been computed before.

Anomalies  associated with correlation functions of conserved currents can be analysed by coupling the currents to external sources, which in our case form an ${\cal N}=1$ superconformal multiplet. As such, the anomaly we discuss here could be related to existing superspace results on anomaly candidates for $D=4$, $\cn=1$ supergravity theories \cite{Bonora:1984pn,Buchbinder:1986im,Brandt:1993vd, Brandt:1996au, Bonora:2013rta} (in particular, in type II anomalies in \cite{Brandt:1996au}), though we emphasise that in our case the supergravity fields are external and thus non-dynamical (off-shell). 


A supersymmetry anomaly appears in super Yang-Mills (SYM) theory in the WZ gauge when there are gauge anomalies \cite{Itoyama:1985qi} (see also \cite{Piguet:1984aa,Guadagnini:1985ea, Zumino:1985vr}). This anomaly is easy to understand: in the WZ gauge, supersymmetry transformations require a compensating gauge transformation and this transfers the anomaly from the gauge sector to supersymmetry. When the SYM theory is consistent at the quantum level (i.e. the gauge anomalies cancel) then supersymmetry is also non-anomalous.  A supersymmetry anomaly appears in theories with gravitational anomalies \cite{Howe:1985uy, Tanii:1985wy, Itoyama:1985ni}, as one may anticipate based on the fact that the energy momentum tensor and the supercurrent are part of the same supermultiplet. Indeed this supersymmetry anomaly sits in the same multiplet as the gravitational anomaly.

Here we will discuss a supersymmetry anomaly in consistent QFTs (no gauge anomalies) which have a conserved energy momentum tensor. We also emphasise that we are concerned with local anomalies, not with beta functions.


\begin{table*}
 \centering
 \begin{tabular}{|l|}
 \hline\\
 \,\,$ \d e^a_\m=\;\x^\l\pa_\l e^a_\m+e^a_\l\pa_\m\x^\l-\l^a{}_b e^b_\m+\s e^a_\m-\frac12\lbar\j_\m\g^a \varepsilon,  \quad
 \d\j_\m=\;\x^\l\pa_\l\j_\m+\j_\l\pa_\m\x^\l-\frac14\l_{ab}\g^{ab}\j_\m+\frac12\s\j_\m+D_\m\ve-\g_\m\h- i\g^5\th\j_\m,  $ \\\\ 
 \,\,$\d A_\m=\;\x^\l\pa_\l A_\m+A_\l\pa_\m\x^\l+\frac{3i}{4}\lbar\f_\m\g^5\ve-\frac{3i}{4}\lbar\j_\m\g^5\h+\pa_\m\th, \qquad \f_\m\equiv\; \frac13\g^\n\big(D_\n\j_\m-D_\m\j_\n-\frac{i}{2}\g^5 \e_{\n\m}{}^{\r\s}D_\r\j_\s\big)$\\\\
 \,\,$[\d_\ve,\d_{\ve'}]=\d_\x+\d_\l+\d_\th,\qquad \x^\m=\frac12\lbar\ve'\g^\m\ve,\quad \l^a{}_b=-\frac12(\lbar\ve'\g^\n\ve)\;\o_\n{}^a{}_b,\quad \th=-\frac12(\lbar\ve'\g^\n\ve)A_\n$\\\\
\,\,$ [\d_\ve,\d_\h]=\d_\s+\d_\l+\d_\th,\qquad \s=\frac12\lbar\ve\h,\quad\l^a{}_b=-\frac12\lbar\ve\g^a{}_b\h,\quad \th=-\frac{3i}{4}\lbar\ve\g^5\h$\\\\
 \hline
 \end{tabular}
 \caption{Transformation rules of the current sources and their algebra, to leading order in the gravitino. All other commutators vanish, except for that of two diffeomorphisms and two local Lorentz transformations, which take a standard form.}
 \label{Local}
 \end{table*}

{\em Holographic anomalies.}---  The anomaly we discuss here was first 
computed holographically \cite{Papadimitriou:2017kzw}.  In holography, 
given a bulk action, one can use holographic renormalisation 
\cite{Henningson:1998gx, deHaro:2000vlm} to compute the Ward identities
and anomalies of the dual QFT. AdS/CFT relates ${\cal N} =1$ SCFT in 
four dimensions to ${\cal N} =2$ gauged supergravity in five dimensions. 
Starting from gauged supergravity in an asymptotically locally AdS$_5$ 
spacetime and turning on sources for all superconformal currents one can 
compute the complete set of superconformal anomalies. This computation 
is available for holographic CFTs, which in particular means that 
the central charges should satisfy $a=c$ as $N \to \infty$ 
\cite{Henningson:1998gx}.

Early attempts to compute the supertrace Ward identity can be found in 
\cite{Chaichian:2003kr,Chaichian:2003wm} but these missed contributions
to the anomaly involving the R-symmetry current and the Ricci tensor. 
Following the work of Pestun \cite{Pestun:2007rz}, there was renewed 
interest in supersymmetric theories on curved spacetimes and their 
holographic duals. The holographic anomalies for bosonic currents were 
computed in \cite{Cassani:2013dba}, reproducing (and correcting) known 
field theory results \cite{Anselmi:1997am}. The full superconformal 
anomalies for the $\cn=1$ current multiplet were computed holographically 
in \cite{Papadimitriou:2017kzw}, while \cite{An:2017ihs} obtained the 
superconformal anomalies in the presence of local supersymmetric scalar
couplings. An analogous holographic computation relevant to two-dimensional SCFTs was reported in
\cite{An:2018roi}.
 
The holographic results leave open the possibility that the anomaly is special to holographic theories at strong coupling. In this Letter we show that this is not the case. One could have anticipated the anomaly based on the structure of the supersymmetric variation of the supercurrent,
which is of the schematic form
$\delta \cq^\mu \sim \gamma_\nu \ct^{\mu \nu} \varepsilon+ C^{\mu \nu \rho} \partial_\nu \cj_\rho \varepsilon$,
where  $C^{\mu \nu \rho}$ is a tensor constructed from gamma matrices and the  metric.
The Ward identity for the 4-point function involving two supercurrents and two R-currents would then involve terms of the form
\begin{align} \label{susy1}
&\partial^{x_1}_{\mu} \langle \cq^\mu(x_1) \bar{\cq^\nu}(x_2) \cj^\kappa(x_3) \cj^\lambda (x_4) \rangle \\
& \qquad \sim \delta(x_1-x_2) \langle \delta \bar{\cq}^\nu(x_2) \cj^\kappa(x_3) \cj^\lambda (x_4) \rangle + \cdots \nonumber\,,
\end{align} 
where the dots denote additional terms (the exact Ward identity is given (\ref{WI_4pt})).
Using the variation of the supercurrent we find that the r.h.s. contains the 3-point function of three R-currents, which is anomalous, and correspondingly one may anticipate (\ref{susy1}) will be anomalous.
Similarly, the same 4-point function but with one of the R-currents replaced by an energy momentum tensor is expected to be anomalous,  since 
$\langle \cj \ct \ct \rangle$ is anomalous. To determine whether an anomaly appears or not we need to carry out the computation explicitly. Before we turn to this, we discuss the consistency condition that the anomalies must satisfy.

{\em Wess-Zumino consistency.}--- Let $e_\mu^a$, $A_\m$ and $\j_\m$ denote the sources (vierbein, gauge field and gravitino) that couple to the superconformal currents and 
$\mathscr{W}[e,A,\j]$ be the generating functional of connected graphs. We define the currents in the presence of sources (as usual) by
\be\label{Majorana-currents}
\ct^\m_a=e^{-1}\frac{\d\mathscr{W}}{\d e^a_\m},\quad
\cj^\m=e^{-1}\frac{\d\mathscr{W}}{\d A_\m},\quad
\cq^\m=e^{-1}\frac{\d\mathscr{W}}{\d\lbar\j_\m}\,,
\ee
where $e\equiv\det(e_\m^a)$. In the presence of anomalies
\be\label{W-anomalies-CS}
\d_{i} \mathscr{W}=\int d^4x \;e \;\epsilon_i\ca_{i}\,,
\ee
where $\delta_{i}$ denotes the superconformal transformations, $\epsilon_i$ are the (local) parameters of the transformations and $\ca_i$ are the corresponding anomalies. The variations form an algebra, $[\delta_{i}, \delta_j] = f^k_{ij} \delta_k$, and using this in
(\ref{W-anomalies-CS}) we obtain the WZ consistency condition
\be \label{WZ}
\int d^4x \left(\d_i (e\;\epsilon_j \ca_{j}) - \d_j(e\;\epsilon_i  \ca_{i}) - f_{ij}^k e\;\epsilon_k \ca_k \right)=0\,.
\ee
The transformation rules and the local algebra they satisfy are derived in \cite{Papadimitriou:2019gel} and are given in Table \ref{Local}.

 \begin{table*}
 \centering
 \begin{tabular}{|ll|l|}
 \hline&&\\
 \,\,\,$e^a_\m\ct^\m_a+\frac12\lbar\j_\m \cq^\m=\ca_W$, &
 \,\,\,$\nabla_\m \cj^\m+i\lbar\j_\m\g^5 \cq^\m=\ca_R$ & \,\,\,Weyl square: \,\,\,\\&& \,\,\,\,\rule{.0cm}{.3cm}$W^2 \equiv W_{\m\n\r\s}W^{\m\n\r\s}$\,\,\,\\
 \,\,\,$D_\m \cq^\m-\frac12\g^a\j_\m \ct^\m_a-\frac{3i}{4}\g^5\f_\m\cj^\m=\ca_Q$, &
 \,\,\,$\g_\m \cq^\m-\frac{3i}{4}\g^5\j_\m\cj^\m=\ca_{S}$ & \,\rule{.0cm}{.5cm} Euler density:\\&& \,\,\,\rule{.0cm}{.3cm}$E=R_{\m\n\r\s}R^{\m\n\r\s}-4R_{\m\n}R^{\m\n}+R^2$\,\\
 \cline{1-2}\multicolumn{2}{|l|}{}& \,\,\,\rule{.0cm}{.5cm}Pontryagin density:\\
 \,\,\,$\ca_W=\frac{c}{16\p^2}\big(W^2-\frac{8}{3}F^2\big)-\frac{a}{16\p^2} E+\co(\j^2)$, &
 \,\,\,$\ca_R=\frac{(5a-3c)}{27\p^2}\;\wt FF+\frac{(c-a)}{24\p^2}\cp$ & \,\,\vspace{-.1cm} \rule{.0cm}{.35cm}$\cp\equiv\wt R^{\m\n\r\s}R_{\m\n\r\s}$\\\multicolumn{2}{|l|}{} & \,\,\rule{.0cm}{.5cm} $\wt R_{\m\n\r\s}\equiv\frac12\e_{\m\n}{}^{\k\l}R_{\k\l\r\s}$ \\
 \multicolumn{2}{|l|}{\,\,\,$\ca_Q=-\frac{(5a-3c)i}{9\p^2}\wt F^{\m\n}A_\m\g^5\f_\n
 +\frac{(a-c)}{6\p^2} \big(\nabla_\m (A_\r \wt R^{\r\s\m\n})\g_{(\n}\j_{\s)}-\frac{1}{4}F_{\m\n} \wt R^{\m\n\r\s} \g_\r\j_\s\big)+\co(\j^3)$\,} & \,\,\,\rule{.0cm}{.5cm}Schouten tensor:  \\\multicolumn{2}{|l|}{} & \,\,\rule{.0cm}{.3cm} $P_{\m\n}\equiv\frac12\big(R_{\m\n}-\frac16Rg_{\m\n}\big)$ \\
 \multicolumn{2}{|l|}{\,\,\,$\ca_{S}=\frac{(5a-3c)}{6\p^2}\wt F^{\m\n}\big(D_\m-\frac{2i}{3}A_\m\g^5)\j_{\n}+\frac{ic}{6\p^2} F^{\m\n}\big(\g_{\m}{}^{[\s}\d_{\n}^{\r]}-\d_{\m}^{[\s}\d_{\n}^{\r]}\big)\g^5D_\r\j_\s $} & \,\vspace{-.1cm}\rule{.0cm}{.6cm} U(1)$_R$ field strengths: \\
 \multicolumn{2}{|l|}{\hspace{1.1cm}\rule{.0cm}{0.5cm}$
 +\frac{3(2a-c)}{4\p^2}P_{\m\n}g^{\m[\n}\g^{\r\s]}D_\r\j_\s+\frac{(a-c)}{8\p^2}\big(R^{\m\n\r\s}\g_{\m\n}-\frac12Rg_{\m\n}g^{\m[\n}\g^{\r\s]}\big)D_\r\j_\s+\co(\j^3)$} & \vspace{-.3cm} \,\, $\wt F_{\m\n}\equiv\frac12 \e_{\m\n}{}^{\r\s}F_{\r\s}$ $F^2\equiv F_{\m\n}F^{\m\n}$ \\\multicolumn{2}{|l|}{} & \\
 \multicolumn{2}{|l|}{} & \vspace{-.2cm}\,\, $F\wt F\equiv F_{\m\n}\wt F^{\m\n}$\\&&\\
\hline
 \end{tabular}
 \caption{Anomalous Ward identities and corresponding anomalies \cite{Papadimitriou:2019gel}. ($D_\m\j_\n\equiv (\pa_\m+\frac14\o_\m{}^{ab}(e,\j)\g_{ab}+i\g^5A_\m)\j_\n-\G^\r_{\m\n}\j_\r$ with $\o_\m{}^{ab}(e,\j)\equiv \o_\m{}^{ab}(e)+\frac14\big(\lbar\j_a\g_\m\j_b+\lbar\j_\m\g_a\j_b-\lbar\j_\m\g_b\j_a\big)$; $\nabla_\mu$ is the Levi-Civita connection; $\phi_\mu$ is defined in Table \ref{Local}.)}
 \label{WIDs}
 \end{table*}
Assuming the R-symmetry current has the standard triangle anomalies (i.e. assuming the from of
$\ca_R$ in Table \ref{WIDs}) the WZ consistency conditions (\ref{WZ}) may be viewed as equations to determine the remaining anomalies. This computation is presented in \cite{Papadimitriou:2019gel} and the results are summarised in Table \ref{WIDs}. Note in particular that all anomalies are given in terms of the central changes $a$ and $c$. The anomalies of the bosonic currents are in agreement with the results derived in \cite{Anselmi:1997am,Cassani:2013dba}. The supersymmetry anomaly $\ca_Q$ that we discuss here is related to the R-symmetry anomaly $\ca_R$ through the same descent equation that relates the supersymmetry anomaly discussed in \cite{Itoyama:1985qi} to the corresponding gauge anomaly. However, as noted earlier, there are important differences in the physics (in  \cite{Itoyama:1985qi} the gauge anomalies must vanish for consistency of the model, while this is not so for the R-anomalies relevant for us), as well as in the context (the WZ conditions discussed in \cite{Itoyama:1985qi} are for a vector multiplet in flat space, while the anomalies in Table \ref{WIDs} are those of $\cn=1$ conformal supergravity \cite{Papadimitriou:2019gel}). 


Here we only discuss one of the WZ equations: the one obtained by considering the commutator of  R-symmetry (with parameter $\theta$) with Q-supersymmetry (with parameter $\varepsilon$):
\be \label{WZ_SUSY}
\int d^4x \big(\d_\varepsilon(e\;\theta \ca_{R}) - \d_\theta(e\;\varepsilon \ca_{Q})\big)=0\,.
\ee
Using the explicit form of  $\ca_{R}$ it is easy to see that $\d_\varepsilon \ca_{R} \neq 0$ and the WZ consistency condition requires that $\ca_{Q} \neq 0$. This argument does not rely on the theory having conformal invariance, and thus we expect any 4d supersymmetric theory with an R-symmetry anomaly to have a corresponding anomaly in the conservation of the supercurrent.\footnote{This expectation has been verified in the followup paper \cite{Papadimitriou:2019yug}.}

One may wonder whether this anomaly can be removed by adding a local counterterm $\mathscr{W}_{\rm ct}$ to the action such that $\mathscr{W}_{\rm ren}=\mathscr{W} +\mathscr{W}_{\rm ct}$ is non-anomalous, i.e. $\d_{\varepsilon} \mathscr{W}_{\rm ren}=0$. Using the commutator of two supersymmetry variations, $[\d_\ve,\d_{\ve'}]$,  given in Table \ref{Local} we find 
\begin{equation} \label{ct}
(\delta_\xi + \delta_\lambda + \delta_\theta) \mathscr{W}_{\rm ren} =0\quad \Rightarrow\quad 
(\delta_\xi + \delta_\lambda) \mathscr{W}_{\rm ren} \neq 0\,,
\end{equation}
since $\delta_\theta \mathscr{W}_{\rm ren} = \ca_R \neq 0$. It  follows that if one wishes to preserve supersymmetry $\mathscr{W}_{\rm ct}$ must break diffeomorphisms and/or local Lorentz transformations.\footnote{Note that since $\ca_R$ is a genuine anomaly it is not possible to set the r.h.s. of the second equation in (\ref{ct}) to zero using a local counterterm. This implies that there are no further local counterterms that can restore diffeomeorphisms and local Lorentz invariance.}
Next, we calculate this anomaly by one-loop computations within a specific model. 

{\em Model.}--- Consider the massless Wess-Zumino action with one complex bosonic field $\phi$ and one Majorana fermionic field $\chi$
\begin{equation}\label{4.12}
 S= - \int d^{4} x \Big( \partial_{\mu}\phi\partial^{\mu}\phi^{*} + \frac{1}{2}\bar{\chi}\slashed{\partial}\chi \Big).
 \end{equation}
 The conserved currents are given in Table \ref{flat-currents}. We have included improvement terms so that classically ${\cal T}^\mu_\mu=0, \gamma^\mu \cq_\mu=0$ and we are dealing with an ${\cal N}=1$ SCFT.
 
 \begin{table*}
 \begin{tabular}{|l|}
 \hline\\
 \,\,\,$\ct^\m_{a}=(\h^{\m\r}\h^\s_a+\h^{\m\s}\h^\r_a-\h^{\m}_\a\h^{\r\s})\pa_\r\f^*\pa_\s\f-\frac{1}{3}\big(\pa^\m\pa_a-\h^\m_a\pa^2\big)(\f^*\f)+\frac14\lbar\c(\g^\m\pa_a+\g_a\pa^\m)\c$\,\,\,\\\\
 \,\,\,$\cj^\m=\frac{2i}{3}\big(\f^*\pa^\m\f-\f \pa^\m\f^*+\frac{1}{4}\lbar\c\g^\m\g^5\c\big)$\\\\
 \,\,\,$\cq^\m=\frac{1}{\sqrt{2}}(\slashed\pa\f \g^\m\c_R+\slashed\pa\f^*\g^\m\c_L)+\frac{\sqrt{2}}{3}\g^{\m\n}\pa_\n(\f \c_R+\f^* \c_L), \quad \c_L\equiv\frac12(1+\g^5)\c, \ \c_R\equiv\frac12(1-\g^5)\c.$\,\,\\\\
 \hline
 \end{tabular}
 \caption{The (on-shell) energy-momentum tensor, ${\ct}^\mu_a$, the R-symmetry current, $\cj^\mu$, and the supersymmetry current, $\cq^\mu$, for the massless superconformal WZ model in flat space. }
 \label{flat-currents}
 \end{table*}

From the form of the anomaly $\ca_Q$ in Table \ref{WIDs} follows that the first anomalous contribution in flat space correlators appears in 4-point functions involving two supercurrents and either two R-currents or an R-current and an energy momentum tensor. Here we discuss the former, referring to \cite{followup} for a detailed account of both cases. 

Since we seek to investigate the possibility of a supersymmetry anomaly, we should not assume the existence of a supersymmetric regulator: the one-loop computation should not be done in superspace.\footnote{On the other hand, the form of anomalies respects the symmetries they break and thus one may use superspace to analyse possible anomaly candidates.}
We will instead do the computation in components and use the same regulator as in the original triangle anomaly computation, namely momentum cut-off \cite{Adler:1969gk,Bell:1969ts}.
 We will consider the 4-point correlation function
\be\label{4pt-fn}
\left \langle \cq^{\mu}(x_{1})\bar{\cq}^{\nu}(x_{2})\cj^{\kappa}\left(x_{3}\right)\cj^{\lambda}\left(x_{4}\right) \right\rangle\,. 
\ee
Standard path integral manipulations show that this correlator classically satisfies the following Ward identity: 
\begin{eqnarray} \label{WI_4pt}
 &&\nonumber -i\partial^{x_1}_{\mu}\left\langle \cq^{\mu}_1 \bar{\cq}^{\nu}_2 \cj^{\kappa}_3 \cj^{\lambda}_ 4\right\rangle = \delta^{(4)}(x_{12}) 
 \left\langle \delta\bar{\cq}^{\nu}_1 \cj^{\kappa}_3 \cj^{\lambda}_4 \right\rangle  \\
  &&\nonumber  
 + \Big[
 \delta^{(4)}(x_{13})\left\langle \delta \cj^{\kappa}_1 \bar{\cq}^{\nu}_2 \cj^{\lambda}_4 \right\rangle
  - \partial^{x_1}_{\rho} \left(\delta^{(4)}(x_{13}) \left\langle \delta \cj'^{\rho \kappa}_1 \bar{\cq}^{\nu}_2 \cj^{\lambda}_4 
 \right\rangle \right)  \\
 &&   
  +(3, \kappa) \leftrightarrow (4, \lambda)\Big]
 - \partial^{x_1}_{\rho} \left(\delta^{(4)}(x_{12}) \left\langle \delta\bar{\cq}^{\prime \nu\rho}_1 \cj^{\kappa}_3 \cj^{\lambda}_4 \right\rangle \right), 
 \end{eqnarray}
where we have used the shorthand notation $\cq^\mu(x_i) \equiv \cq^\mu_i$, etc., $x_{ij}\equiv x_i-x_j$, and the contributions on the r.h.s. are expressed in terms of the supersymmetry variations of the currents: 
$
\d_\varepsilon \cq^\mu {=} \varepsilon \delta \cq^{\mu} {+} \partial_{\nu} \varepsilon \delta \cq'^{\mu \nu}$ and idem for $\cj^\mu$.
A similar Ward identity follows from R-invariance.

\begin{figure}[h]
\begin{center}
\includegraphics[width=0.4 \textwidth]{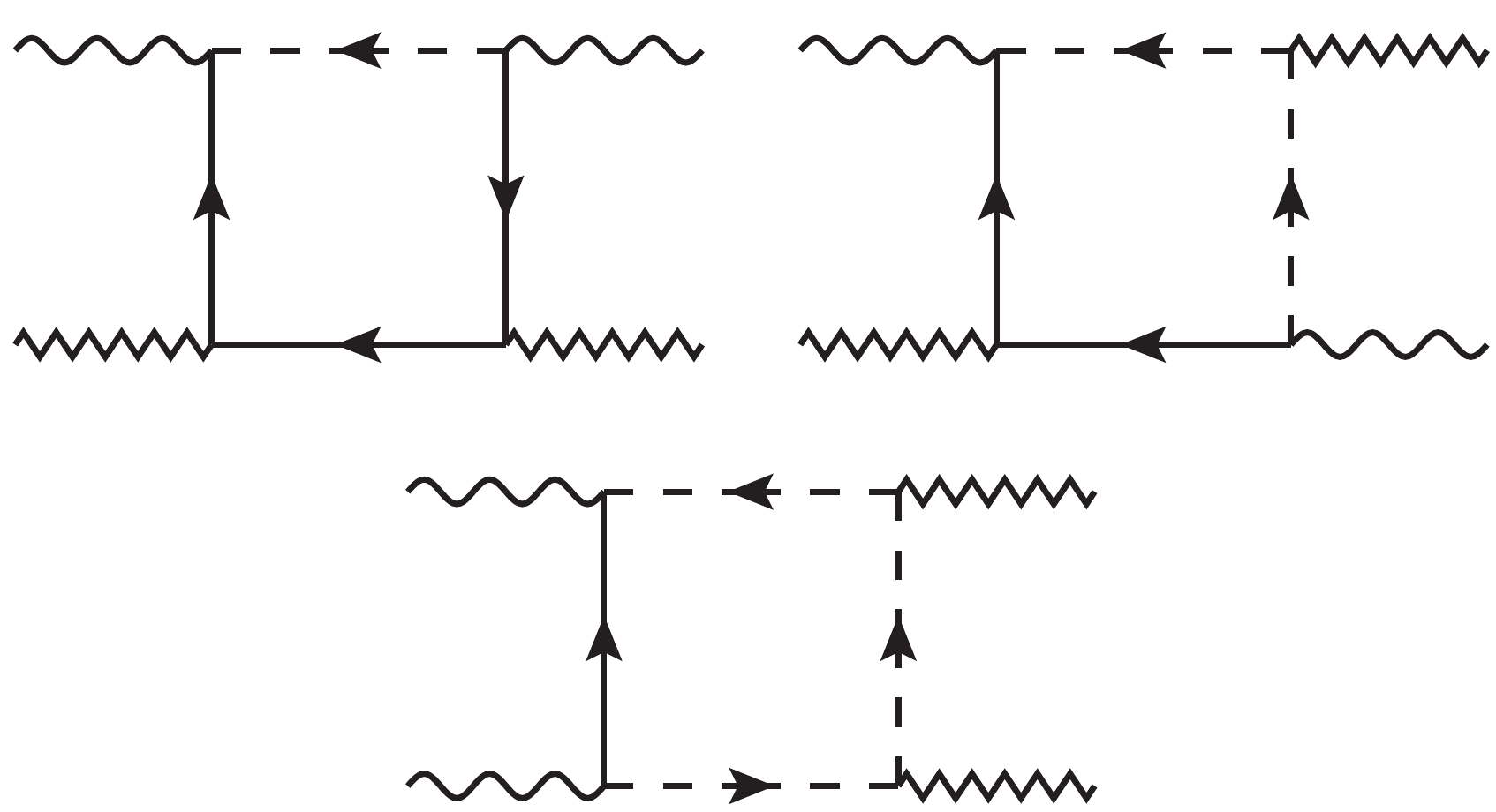}
\caption{\label{fig:1} Box diagrams contributing to the 4-point correlation function \eqref{4pt-fn}.
Zig-zag lines denote R-currents; wavy lines denote supersymmetry currents; straight lines denote fermionic progagators and dashed lines denote bosonic propagators.}
\end{center}
\end{figure}

{\em One-loop computation.}--- We now compute \eqref{WI_4pt}. Since the theory is free the complete computation is one-loop. The 4-point function receives contributions from three classes of Feynman box diagrams,
shown in Figure~\ref{fig:1}; this computation is straightforward but tedious.

One may verify that (9), as well as the corresponding R-symmetry Ward identity, are (naively) satisfied by a simple shift of the loop momentum, much the same way as the triangle Ward identity is naively satisfied. Again in parallel with the triangle anomaly,  (part of) the one-loop contributions to the 4-point function are superficially linearly divergent. This implies that there is a momentum routing ambiguity when using a momentum cut-off regulator (see for example Jackiw's lectures in \cite{Treiman:1986ep}). 

We proceed by taking the $\partial^{x_3}_\kappa$ of \eqref{WI_4pt} and subtracting from it the $\partial^{x_1}_\mu$ derivative of the corresponding R-symmetry Ward identity. By construction, the 4-point functions cancel and one is left with an identity involving 3-point functions only (namely the terms appearing on the r.h.s. of the Ward identities). Had these 3-point functions been non-anomalous, this would be an identity. However, the 3-point functions involve the anomalous $\langle \cj \cj \cj\rangle$ correlator and this implies that either (\ref{WI_4pt}) or the corresponding R-symmetry Ward identity should be anomalous. Assuming the form of the bosonic Ward identities is standard (i.e. given by the expressions in Table \ref{WIDs}) the R-symmetry 4-point function Ward identity is not anomalous and therefore the supersymmetry Ward identity is anomalous. This computation is the counterpart of
\eqref{WZ_SUSY} but now in terms of Feynman diagrams.

One can then show that there is a momentum routing such that 1) the triangle R-symmetry anomaly is reproduced; 2) the 4-point R-symmetry Ward identity is non-anomalous and 3) the supersymmetry Ward identity is anomalous, with the anomaly given in Table \ref{WIDs} and with $c=2 a=1/24$, which are the values in our model. In addition, upon taking the gamma trace of the same 4-point function, $\gamma_\mu \left\langle \cq^{\mu}\bar{\cq}^{\nu} \cj^{\kappa} \cj^{\lambda} \right\rangle$, one automatically reproduces the $\ca_S$ anomaly given in Table \ref{WIDs}. 

In general, changing the momentum routing one may move the anomaly from one conserved current to another. This would be equivalent to adding local finite counterterms and as argued earlier there is no choice of such counterterms that would remove the supersymmetry anomaly while preserving diffeomorphisms/local Lorentz transformations.

It is also straightforward to check that the same anomaly is present in the massive WZ model as well.
As in the case of standard triangle anomalies, adding a mass term modifies the Ward identities but the anomaly remains the same. This is as expected since the anomaly arises from the UV behaviour of  
Feynman diagrams and the parts of the loop computation that give rise to the anomaly remain the same.

{\em Implications of the anomaly.}---
Let us conclude with a few comments about the implications of this anomaly. As mentioned earlier, an important consequence is that a SQFT with such a supersymmetry anomaly cannot be coupled to dynamical supergravity.\footnote{The anomalous R-symmetry alone implies that coupling to a supergravity that gauges the R-symmetry is inconsistent. Here we see that couplings to supergravity that do not gauge the R-symmetry are also inconsistent.} In the context of supersymmetric model building, one does not usually work with theories with an R-symmetry, anomalous or non-anomalous; non-anomalous R-symmetry is not compatible with gaugino masses (see \cite{Drees:2004jm}). More generally, one does not expect a theory with continuous symmetry to emerge from a consistent quantum theory of gravity, such as string theory. However, such  models may be considered in bottom-up approaches (see \cite{Pallis:2018xmt} for a recent example). Similar comments apply to bottom-up string cosmology models. This anomaly also affects supersymmetric localisation computations, as has already been noted in \cite{Papadimitriou:2017kzw,An:2017ihs,An:2018roi,Papadimitriou:2019gel}.  However, it is possible that a suitable non-covariant local counterterm\footnote{For theories with $a=c$ such a counterterm evaluated on supersymmetric backgrounds of the form $S^1\times M_3$, with $M_3$ a Seifert manifold, should agree with the counterterm used in \cite{Genolini:2016ecx}.} may cancel the rigid supersymmetry anomaly at the expense of breaking certain diffeomorphisms on a given supersymmetric background. From a more formal perspective, it would be interesting to explore how the supersymmetry anomaly is captured in index theorems. It would also be interesting to investigate the existence of such an anomaly in other dimensions and/or extended supersymmetry.

{\bf Note Added:} While this paper was finalised, a related work \cite{An:2019zok} appeared on the arXiv.

\begin{acknowledgments}
{\em Acknowledgments.}--- We would like to thank Benjamin Assel, Roberto Auzzi, Friedmann Brandt, Loriano Bonora, Davide Cassani, Cyril Closset, Camillo Imbimbo, Manthos Karydas, Heeyeon Kim, Zohar Komargodski, Dario Martelli, Sunil Mukhi, Sameer Murthy, Parameswaran Nair, Dario Rosa, Stanislav Schmidt, Ashoke Sen and Peter West for illuminating discussions and email correspondence. 
KS and MMT are supported in part by the Science and Technology Facilities Council (Consolidated Grant ``Exploring the Limits of the Standard Model and Beyond''). This research was supported in part by the National Science Foundation under Grant No. NSF PHY-1748958 and this project has received funding/support from the European Union's Horizon 2020 research and innovation programme under the Marie Sklodowska-Curie grant agreement No 690575. IP would like to thank the University of Southampton, King's College London, and the International Center for Theoretical Physics in Trieste for hospitality and partial financial support during the completion of this work. MMT would like to thank the Kavli Institute for the Physics and Mathematics of the Universe for hospitality during the completion of this work. 
\end{acknowledgments}

\bibliographystyle{apsrev4-1}
\bibliography{susy-anomalies}

\end{document}